\documentclass{PoS}
\usepackage[utf8]{inputenc}
\usepackage{graphicx}
\usepackage{amsmath}
\usepackage{amssymb}
\usepackage{booktabs}
\usepackage{graphics}

\newcommand{\MKK}{M_{\rm KK}}
\newcommand{\uKK}{u_{\rm KK}}

\newcommand{\be}{\begin{equation}}
\newcommand{\ee}{\end{equation}}
\newcommand{\bea}{\begin{eqnarray}}
\newcommand{\eea}{\end{eqnarray}}

\def\PDG{\cite{Agashe:2014kda}}
\def\BPR{\cite{Brunner:2015oqa}}
\def\BR{\cite{1504.05815}}

\title{Glueball Decay in the Witten-Sakai-Sugimoto Model and Finite Quark Masses}

\ShortTitle{Glueball Decay in the Witten-Sakai-Sugimoto model}

\author{\speaker{Frederic Brünner} and Anton Rebhan\\
       Institute f\"ur Theoretische Physik, Technische Universit\"at Wien,\\ Wiedner Hauptstra{\ss}e 8-10, A-1040 Vienna, Austria\\
        E-mail: \email{bruenner@hep.itp.tuwien.ac.at}, \email{rebhana@hep.itp.tuwien.ac.at}}

\abstract{We discuss recent results on the calculation of glueball decay rates in the Witten-Sakai-Sugimoto model, which favor the $f_0(1710)$ meson as a glueball candidate. 
The flavor asymmetric decay of $f_0(1710)$ is frequently attributed to a putative chiral suppression in glueball decays, which
is however questionable in view of the large constituent quark masses induced by chiral symmetry breaking.
We find that this can be explained by
what we call nonchiral enhancement when finite quark masses are included in the holographic model, with good quantitative
agreement with experimental data for $f_0(1710)$. Assuming the latter to indeed be a nearly pure glueball, the model makes essentially parameter-free and thus falsifiable predictions for its decay rates involving vector mesons and an upper limit on the $\eta\eta'$ decay rate.}

\FullConference{The 8th International Workshop on Chiral Dynamics, CD2015 ***\\
		29 June 2015 - 03 July 2015\\
		Pisa,Italy}

\begin{document}

\section{Introduction}

The existence of glueballs, 
color-neutral bound states of gluons, is a particularly intriguing prediction of
quantum chromodynamics (QCD) dating back to the early 1970's
\cite{Fritzsch:1972jv,Fritzsch:1975tx,Jaffe:1975fd}.
Lattice QCD \cite{%Bali:1993fb,Morningstar:1999rf,
Gregory:2012hu}
predicts the lightest glueball to be a scalar with mass
in the range 1.5-1.8 GeV, but does not pin down sufficiently
its potential mixing with scalar quark-antiquark states and its decay pattern.
There exist a large number of phenomenological studies which
model the three isoscalar mesons $f_0(1370)$, 
$f_0(1500)$, and $f_0(1710)$ as being made up from mixtures of
$u\bar u+d\bar d$, $s\bar s$, and a scalar glueball,
alternatingly locating the glueball predominantly in
$f_0(1500)$ or $f_0(1710)$ 
\cite{Amsler:1995td,Lee:1999kv,Close:2001ga,Amsler:2004ps,Close:2005vf,Giacosa:2005zt,Albaladejo:2008qa,Mathieu:2008me,Janowski:2011gt,Janowski:2014ppa,Cheng:2015iaa,Close:2015rza,Frere:2015xxa}.
Models which identify $f_0(1500)$ as dominated by the glueball component typically have
rather large mixing with $q\bar q$ states, while several of the models which favor
$f_0(1710)$ as a glueball do so with comparatively little admixture of $q\bar q$.

The fact that $f_0(1710)$ decays preferentially into kaons and $\eta$ mesons and less into
pions, which goes contrary to the expectation of a flavor-blindness of glueball decays,
has been attributed to
a mechanism termed ``chiral suppression'' \cite{Carlson:1980kh,Sexton:1995kd,Chanowitz:2005du}
according to which the decay amplitudes of a scalar glueball should be proportional to
quark masses. At least the perturbative arguments in its favor appear questionable
\cite{Frere:2015xxa} in view of the large constituent quark masses brought about by
chiral symmetry breaking.

In this paper we shall review our recent work on scalar glueball decay using the top-down holographic
Witten-Sakai-Sugimoto model \BPR, which is flavor-symmetric, and its extension to finite quark masses which break SU(3)$_f$
\cite{1504.05815,1510.07605}, where we have shown the possibility of what we called
nonchiral enhancement in scalar glueball decay. The observed flavor asymmetries of the decays of $f_0(1710)$
into two pseudoscalars turn out to be reproduced well in coincidence with a very small branching ratio
$G\to\eta\eta'$.

\section{The Witten-Sakai-Sugimoto model and meson decay rates}

The best studied string-theoretic realization of gauge/gravity duality relates
ten-dimensional type IIB supergravity on $AdS_5\times S^5$ to strongly coupled 
four-dimensional maximally supersymmetric Yang-Mills theory
in the large-$N_c$ limit. 
This has been widely used to study strongly coupled nonabelian gauge theories in the deconfined phase, where
supersymmetry is broken by nonzero temperature.
Due to its high amount of symmetries, this correspondence is however not suited to study low-energy QCD. Already in 1998 Witten \cite{Witten:1998zw} introduced a nonconformal version of the correspondence 
based on type IIA supergravity and five-dimensional super-Yang-Mills theory, compactified to give
nonsupersymmetric four-dimensional Yang-Mills theory below a certain compatification scale $\MKK$.

A D-brane construction which introduces $N_f\ll N_c$ chiral quarks was found
by Sakai and Sugimoto \cite{Sakai:2004cn,Sakai:2005yt} and shown to reproduce various
features of low-energy QCD. In particular, chiral symmetry breaking $\mathrm{U}(N_f)_L\times\mathrm{U}(N_f)_R
\to\mathrm{U}(N_f)_V$ is realized purely geometrically, and an effective field theory
involving the associated Nambu-Goldstone bosons, vector and axial vector
mesons can be derived with all couplings fixed by just two parameters,
the mass scale $\MKK$ and the 't Hooft coupling $\lambda$ at this scale.
It also includes the correct Wess-Zumino-Witten term and a well-defined Witten-Veneziano mass term for the $\eta'$.

As reviewed in \cite{Rebhan:2014rxa}, this model works also remarkably well quantitatively.
Fixing $\MKK$ by the $\rho$ meson mass and varying $\lambda=16.63\ldots12.55$ such that either the pion decay constant
or the string tension in large-$N_c$ lattice simulations is matched, the decay rate of
the $\rho$ and the $\omega$ meson into pions is predicted to lie in the range \BPR\
\be
\Gamma(\rho\to2\pi)/ m_\rho = 0.1535\dots 0.2034,\quad
\Gamma(\omega\to3\pi)/ m_\omega = 0.0033\ldots 0.0102,
\ee
which nicely include the experimental values, 0.191(1) and
0.0097(1), respectively.

One may therefore entertain the hope that the Witten-Sakai-Sugimoto model could also make useful predictions
for the decay rates of glueballs. 
In fact, the spectrum of glueballs
has been one of the first applications of the Witten model
\cite{hep-th/9805129,Constable:1999gb,Brower:2000rp}, and a first attempt to calculate
the decay rate of the lightest glueball was made in \cite{Hashimoto:2007ze}.
However, the mass of the lowest-lying glueball comes out at only 855 MeV.
In \BPR\ we have revisited these calculations and argued that the lowest mode, which comes
from a graviton mode with an ``exotic'' polarization along the compactification direction, should be
discarded and instead the next-highest, predominantly dilatonic mode should be identified with the lowest-lying glueball 
of QCD. This turns out to have a mass closer to that expected from lattice QCD, to wit 1487 MeV, and also a significantly narrower
width than the exotic mode.

Calculating the decay pattern of the holographic glueball with a mass that would fit to the $f_0(1500)$ meson
however does not match the observed decays of the latter \PDG, which decays mostly into four pions (49.5\%) and two
pions (34.9\%), with kaons and eta mesons suppressed significantly. The holographic result for four-pion decays is
about an order of magnitude too small, 
the result for decay into two (massless) pseudoscalars less than 50\% of the experimental value for two-pion decay,
but more than twice as large as the actual decay into kaons.

However, the holographic result for the glueball mass is only 16\% below the mass of the other glueball candidate, $f_0(1710)$.
It thus seems reasonable to compare the results for the dimensionless quantities $\Gamma/M$ with the experimental data for $f_0(1710)$. Now the holographic result \BPR\ 
\be
\Gamma(G\to\pi\pi)/M=0.009\ldots0.012\quad \text{(vanishing quark masses)}
\ee
appears to much more compatible
with experiment, while the larger rates into kaons and eta mesons would have to be attributed somehow to
the nonnegligible strange quark mass.

\section{Finite quark mass deformation of the Witten-Sakai-Sugimoto model}

In the literature \cite{0708.2839,Dhar:2008um,Niarchos:2010ki,Aharony:2008an,Hashimoto:2008sr,McNees:2008km}, two ways of introducing finite quark masses in the Witten-Sakai-Sugimoto model have been discussed: either by world sheet instantons arising from additional D4 or D6 branes added to the geometry, or by taking into account a nonnormalizable mode in a bifundamental scalar corresponding to the open-string tachyon between D and anti-D branes. It was also suggested that the two approaches were actually two ways of viewing one and the same mechanism. While both approaches correctly reproduce the GOR relations, unfortunately neither 
has been developed to the point of fixing the interactions of glueball modes with the mass term, which has the following nonlocal form
\begin{equation}
 \int d^4x \int_{u_{KK}}^\infty du \; h(u) \mathrm{Tr}\left( \mathcal{T}(u)\mathrm{P}e^{-i\int dz A_z(z,x)}+h.c.\right)
 =\int d^4x \,\mathrm{Tr}\left( U(x)\int_{u_{KK}}^\infty du \; h(u) \mathcal{T}(u)+h.c.\right).
\end{equation}
Here $u$ and $z$ parametrize the holographic direction and are related by $(u/\uKK)^3=1+z^2$, with $\uKK$ denoting the lower end of the cigar-shaped geometry. $\mathcal{T}(u)$ is a bifundamental field 
implementing the quark mass matrix according to
\begin{equation}
\int_{u_{KK}}^\infty du\; h(u)\mathcal{T}(u)\propto\mathcal{M}\equiv\mathrm{diag}(m_u,m_d,m_s)
\end{equation}
and $h(u)$ contains metric and dilaton degrees of freedom, whose fluctuations contain the glueball modes.

\begin{table}
\centering
\resizebox{8cm}{!}{%
\begin{tabular}{lccc}
%\toprule
%$f_0(1710)$ 
& exp.\PDG & WSS massive \BR \\
\toprule
$\frac43 \cdot \Gamma(\pi\pi)/\Gamma(K\bar K)$ &  0.55$+0.15\atop-0.23$  &  0.463 \\
$4 \cdot\Gamma(\eta\eta)/\Gamma(K\bar K)$ &  1.92$\pm0.60$  &  1.12 \\
\end{tabular}%
}
\caption{Comparison between the flavor asymmetries observed in the decay rates of the glueball candidate $f_0(1710)$ and the prediction of the massive Witten-Sakai-Sugimoto model for a pure glueball of that mass.}\label{tab1}
\end{table}

A very similar nonlocal mass term arises already in the massless Witten-Sakai-Sugimoto model from the U(1)$_A$ anomaly,
which can be calculated precisely \cite{Sakai:2004cn}.
The U(1)$_A$ anomaly requires the combination of a bulk Ramond-Ramond 2-form field and the boundary $\eta_0$ field,
\be
\eta_0(x)=\frac{f_\pi}{\sqrt{2N_f}}\int dz \,{\rm Tr}\, A_z(z,x),
\ee
in the action
\begin{equation}
S\propto\int d^{10}x\; \sqrt{-g}|\tilde{F}_2|^2,
\end{equation}
with
\begin{equation}
\tilde{F}_2\propto \left(\theta+\frac{\sqrt{2N_f}}{f_\pi}\eta_0\right)du\wedge dx^4,
\end{equation}
\noindent where $\theta$ is the QCD theta angle, $x^4$ the compactified coordinate of the cigar geometry, and $f_\pi$ the pion decay constant.
Integration over the bulk gives the four-dimensional effective action
\be\label{Seffeta0}
S^{\rm eff.}_{\eta_0}=-\frac12 \int d^4x\, m_{0}^2 \,\eta_0^2 \left( 1-3 d_0 G_D\right)+\ldots
\ee
where $G_D$ is the scalar glueball mode, 
\be\label{d0}
d_0\approx \frac{17.915}{\lambda^{1/2}N_c\MKK},
\ee
and
\be
m_{0}^2=\frac{N_f}{27\pi^2 N_c}\lambda^2\MKK^2
\ee
the Witten-Veneziano mass.
For $N_f=N_c=3$, $\MKK=949$ MeV, and $\lambda$ varied from 16.63 to 12.55 one finds
$m_{0}=967\ldots730$ MeV.

Diagonalizing the mass terms for the entire nonet of pseudoscalar bosons in $U$ gives,
with $\mathcal M={\rm diag}(\hat m,\hat m,m_s)$ fixed such that $m_\pi=140$ MeV and $m_K=497$ MeV:
\bea
&&m_\eta=518\ldots476\; {\rm MeV},\quad 
m_{\eta'}=1077\ldots894\; {\rm MeV},\quad\\
&&\theta_P=-14.4^\circ\ldots-24.2^\circ,
\label{thetaP}
\eea
with $\theta_P$ the octet-singlet mixing angle.
We thus find that the above holographic result for $m_{0}$ is in the right ballpark to approximate
the pseudoscalars of the real world, in fact covering most of the values of $\theta_P$
that are being discussed in the phenomenological literature \cite{Gerard:2004gx,Gerard:2013gya,Pham:2015ina}. 
The central value of the above range of masses and the mixing angle
happens to be very close to an optimal least-square choice of $m_\eta$ and $m_{\eta'}$, with the interesting
prediction $\theta_P\approx-19^\circ$ that is consistent with a determination from
$\Gamma(\eta'\to2\gamma)/\Gamma(\eta\to2\gamma)$, which leads to \PDG\ $\theta_P=(-18\pm2)^\circ$.

\section{Nonchiral enhancement and $\eta\eta'$ decay rate}

If we assume that the couplings of the glueball field to the mass terms of the pseudoscalar fields
are universal, i.e.\ that they involve all the factor $(1-3 d_0 G_D)$ obtained in (\ref{Seffeta0}),
the glueball interactions do not mix $\eta$ and $\eta'$. (In \BR\ we have made this case plausible
by a simplistic calculation that gave a coupling which differed by a mere 4\% from $d_0$.)
In this most symmetric case,
the chiral result for the decay rate $G\to PP$
for a given pseudoscalar of mass $m_P$ turns out to be simply multiplied by the factor
\begin{equation}
\left(1-4\frac{m_P^2}{M^2}\right)^{1/2}\left(1+\alpha\frac{m_P^2}{M^2}\right)^2
\quad\text{with}\;\;
\alpha=4(3d_0/d_1-1)\approx 8.480,
\end{equation}
where $d_1$ is the glueball coupling constant appearing in the chiral $G_D\pi\pi$ term \BPR.
Notice that this (leading-order) result is in fact independent of the two parameters $\lambda$ and $\MKK$ of the
model.

Inserting physical values for $m_K$ and $m_\eta$ and setting $M$ to the mass of the glueball candidate $f_0(1710)$
turns out to reproduce the experimental ratio $\Gamma(\pi\pi)/\Gamma(K\bar K)$ within the experimental error bar, and the
$\Gamma(\eta\eta)/\Gamma(K\bar K)$ within 1.33 standard deviations (Table \ref{tab1}).
This shows that the Witten-Sakai-Sugimoto model with finite quark masses naturally leads to
an enhancement of decays into the heavier pseudoscalars that would be fully compatible with the
interpretation of $f_0(1710)$ as a nearly unmixed glueball.
We prefer to call the mechanism at work here as
a ``nonchiral enhancement'' because the chiral limit worked out in \BPR\ does not lead to a
particular suppression of glueball decay into two Nambu-Goldstone bosons. Glueballs rather receive an additional
coupling from the mass term induced by finite quark masses.

\begin{figure}
\centerline{\includegraphics[width=0.7\textwidth]{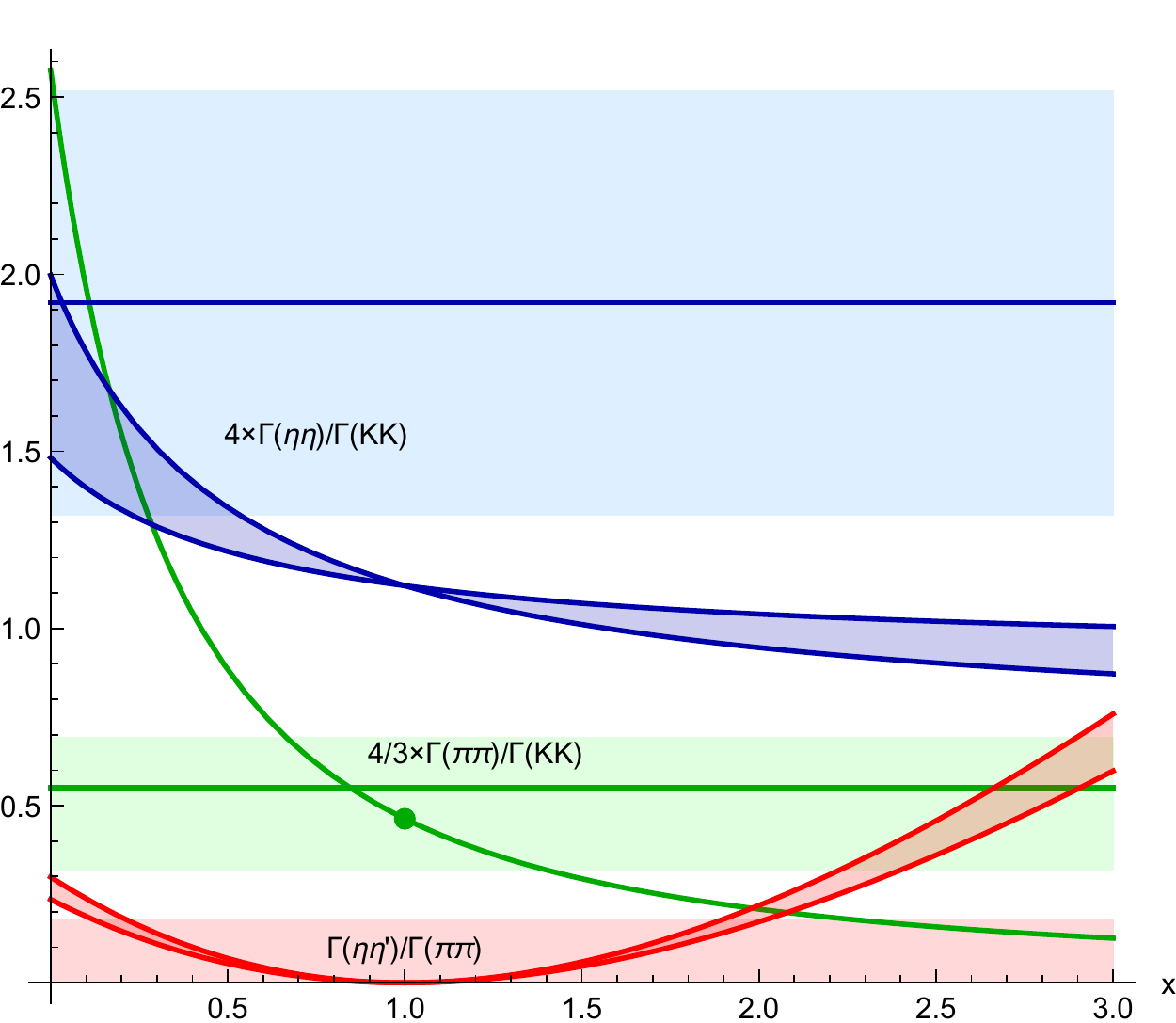}}
\caption{Results \cite{1510.07605} of the massive Witten-Sakai-Sugimoto model on flavor asymmetries in the decay of a glueball of
mass equal to that of $f_0(1710)$ into pairs of pseudoscalar mesons as a function of $x=d_m/d_0$ and with $\lambda$ varied
from 12.55 to 16.63. The case $x=1$ studied in \BR\ and shown in Table~\protect\ref{tab1} is marked by a dot. The light-green and light-blue bands give the current experimental results for
$f_0(1710)$ reported in \PDG, the light-red band corresponds to the upper limit on the $\eta\eta'$ decay rate
of WA102 \cite{Barberis:2000cd}.}
\label{figgrats}
\end{figure}

Since we have not been able to derive this coupling within a string-theoretic top-down construction but just
assumed a particularly symmetric case, one should in fact also consider the consequences of a significantly
different coupling $d_m\not=d_0$ in the GOR mass terms compared to the Witten-Veneziano mass term. 
This most general case leads to a nonzero decay rate $G\to\eta\eta'$ and modified nonchiral enhancements \cite{1510.07605},
such that the still to be determined $\eta\eta'$ decay is restricted by any bounds on the latter and vice versa.
As shown in Fig.~\ref{figgrats}, the current upper limit from WA102 \cite{Barberis:2000cd},
$\Gamma(\eta\eta')/\Gamma(\pi\pi)<0.18$ (red band in Fig.~\ref{figgrats}) is related to a range
of $\Gamma(\pi\pi)/\Gamma(KK)$ that is significantly larger than the current experimental error reported in \PDG.
Under the assumption that the $f_0(1710)$ is a pure glueball, or very close to a pure glueball,
and that the current data on $\Gamma(\pi\pi)/\Gamma(KK)$ hold up, the Witten-Sakai-Sugimoto model
makes the prediction that $\Gamma(\eta\eta')/\Gamma(\pi\pi)\lesssim 0.04$, i.e.\ much smaller than
the current upper bound. (Interestingly enough, the recent phenomenological study in \cite{Frere:2015xxa}
predicts $\eta\eta'$ rates for $f_0(1710)$ that are several times higher than the upper limit reported by WA102.
Hopefully new data, e.g.\ from BESIII, will pin down this decay channel.)

Another prediction of the Witten-Sakai-Sugimoto model, which in fact does not change substantially when
finite quark masses are introduced, is that a pure glueball with a mass of 1.7 GeV should have
a substantial branching ratio into four pions \BPR: $\Gamma(G\to4\pi)/\Gamma(G\to2\pi)\approx 2.5$,
and also into two $\omega$ mesons: $\Gamma(G\to2\omega)/\Gamma(G\to2\pi)\approx 1.1$. While the latter
process has been seen \PDG, the former will hopefully be measured in the near future, e.g.\ by CMS/TOTEM.

To conclude, the Witten-Sakai-Sugimoto model allows one to study flavor asymmetries in glueball decays,
with quantitative results on decay rates that are remarkably close to experimental data on  the glueball candidate $f_0(1710)$,
and it does make a number of falsifiable predictions on still to be measured
decays, if the scalar meson $f_0(1710)$ is indeed
a nearly unmixed glueball.

\begin{acknowledgments}
We thank Denis Parganlija for collaboration on the massless Witten-Sakai-Sugimoto model and many discussions.
This work was supported by the Austrian Science
Fund FWF, project no. P26366, and the FWF doctoral program
Particles \& Interactions, project no. W1252.
\end{acknowledgments}

\bibliographystyle{JHEP}
\bibliography{glueballdecay}

\end{document}